\documentclass[manuscript]{aastex}

\slugcomment{Astrophysical Journal Letters}

\shorttitle{Interstellar Interloper 1I/2017 U1}
\shortauthors{Jewitt et al.}

\begin{document}

\title{Interstellar Interloper 1I/2017 U1: Observations from the \\NOT and WIYN Telescopes}

\author{David Jewitt$^{1,2}$, Jane Luu$^{3,4}$,  Jayadev Rajagopal$^5$, Ralf Kotulla$^6$, Susan Ridgway$^5$, Wilson Liu$^5$ and Thomas Augusteijn$^7$
}

\affil{$^1$ Department of Earth, Planetary and Space Sciences,
UCLA, 
595 Charles Young Drive East, 
Los Angeles, CA 90095-1567\\
$^2$ Dept.~of Physics and Astronomy,
UCLA, 
430 Portola Plaza, Box 951547,
Los Angeles, CA 90095-1547\\
$^3$ Department of Physics and Technology, Arctic University of Tromso, Tromso, Norway \\
$^4$ MIT Lincoln Laboratory, Lexington, MA 02420 \\
$^5$ NOAO, 950 North Cherry Ave., Tucson, AZ 85719 \\
$^6$ Department of Astronomy, University of Wisconsin-Madison, 475 N. Charter St., Madison, WI 53706 \\
$^7$ Nordic Optical Telescope, Apartado 474, 38700 Santa Cruz de La Palma, Santa Cruz de Tenerife, Spain\\
}


\email{jewitt@ucla.edu}


\begin{abstract}
We present observations of the interstellar interloper 1I/2017 U1 ('Oumuamua) taken during its 2017 October flyby of Earth.   The optical colors  B-V = 0.70$\pm$0.06, V-R = 0.45$\pm$0.05, overlap those of the D-type Jovian Trojan asteroids and are incompatible with the ultrared objects which are abundant in the Kuiper belt.  With a mean absolute magnitude  $H_V$ = 22.95 and assuming a geometric albedo $p_V$ = 0.1, we find an average radius of 55 m.  No coma is apparent; we deduce a limit to the dust mass production rate of only $\sim$ 2$\times$10$^{-4}$ kg s$^{-1}$, ruling out the existence of exposed ice covering more than a few m$^2$ of the surface.  Volatiles in this body, if they exist, must lie  beneath an involatile surface mantle $\gtrsim$0.5 m thick, perhaps a product of prolonged cosmic ray processing in the interstellar medium.   The lightcurve range is unusually large at $\sim$2.0$\pm$0.2 magnitudes.  Interpreted as a rotational lightcurve the body has semi-axes $\sim$230 m $\times$ 35 m.   A $\sim$6:1 axis ratio is extreme relative to most small solar system asteroids and suggests that  albedo variations may additionally contribute to the variability.  The lightcurve is consistent with a two-peaked  period  $\sim$8.26 hr but the period is non-unique as a result of aliasing in the data. Except for its unusually elongated shape, 1I/2017 U1 is a physically unremarkable, sub-kilometer, slightly red, rotating   object from another planetary system.  The steady-state population of similar, $\sim$100 m scale interstellar objects inside the orbit of Neptune is $\sim$10$^4$, each with a residence time $\sim$10 yr.

\end{abstract}

\keywords{minor planets, asteroids: general --- minor planets, asteroids: individual (1I/2017 U1) --- comets: general --- ISM: general --- ISM: 1I/2017 U1}

\section{Introduction}

Interstellar interloper 1I/2017 U1 (briefly named C/2017 U1, then A/2017 U1, hereafter ``U1'') was discovered receding from the Sun on UT 2017 October 18.5 (Williams 2017).  The orbit has perihelion distance $q$ = 0.254 AU, eccentricity $e$ = 1.197 and inclination $i$ = 122.6\degr. Perihelion occurred on UT 2017 September 09, five weeks before discovery.  While 337 long-period comets are known with $e >$ 1,  in each case these are Oort cloud comets which have been accelerated above  solar system escape velocity by planetary perturbations and/or reaction forces due to asymmetric outgassing (e.g.~Kr{\'o}likowska \& Dybczy{\'n}ski 2017).  U1 is special because the velocity at infinity is $\sim$25 km s$^{-1}$, far too large to be explained by local perturbations.  It is the first  interstellar interloper observed in the solar system (de la Fuente Marcos and de la Fuente Marcos 2017) and, as such,  presents an opportunity to characterize an object formed elsewhere in our galaxy.

\section{Observations}

\begin{figure}
\epsscale{.9}
\plotone{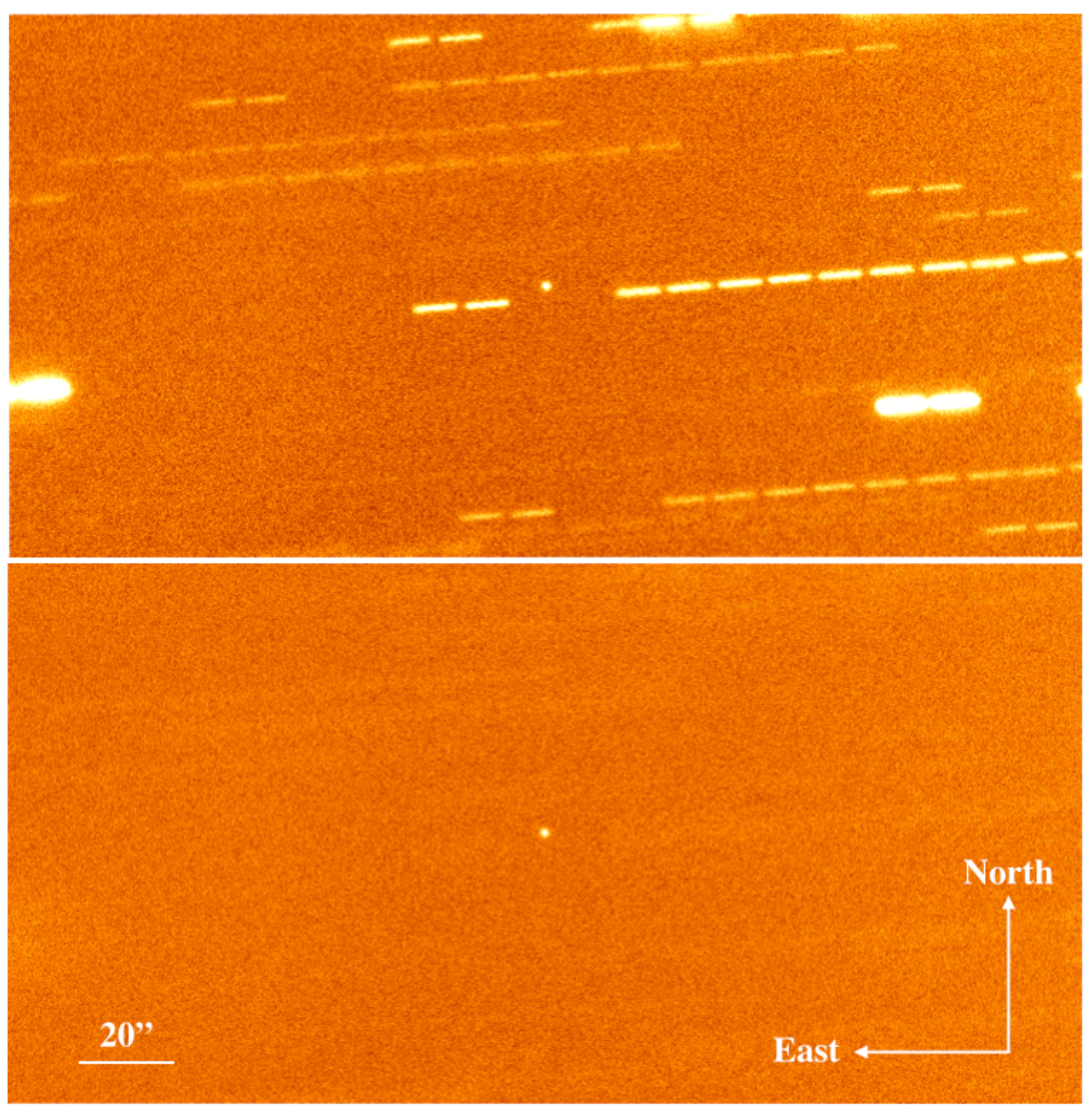}
\caption{Composite images of 1I/2017 U1 taken UT 2017 October 26 at the NOT telescope.  The upper panel shows the average of 12 R-band images, each of  120 s integration.  The median of the same 12 images is shown in the lower panel.  \label{images}}
\end{figure}

Observations were obtained on UT 2017 October 25/26 and 29/30 using the 2.5 meter diameter Nordic Optical Telescope (NOT), located on La Palma, the Canary Islands.  We used the Andalucia Faint Object Spectrograph and Camera (ALFOSC)  optical camera, which is equipped with a  2048$\times$2064 pixel ``e2v Technologies'' charge-coupled device (CCD), an  image scale of 0.214\arcsec~pixel$^{-1}$ and a vignet-limited field of view approximately 6.5\arcmin$\times$6.5\arcmin.   Broadband Bessel B (central wavelength $\lambda_c$ = 4400\AA, full width at half-maximum FWHM = 1000\AA), V ($\lambda_c$ = 5300\AA, FWHM = 800\AA) and R ($\lambda_c$ = 6500\AA, FWHM = 1300\AA) filters were used.   The NOT was tracked at non-sidereal rates to follow U1, while autoguiding using field stars.   We obtained a series of images each of 120 s integration during which time field objects trailed by $\sim$7.5\arcsec.  The images were first bias subtracted and then normalized by a flat field image constructed from images of the sky. The target was readily identified in the flattened images from its position near the center of the image and its stellar appearance while all other field objects were trailed.   Photometric calibration of the data was obtained by reference to standard stars L92 355, L92 430 and PG +0231 051.  Measurements of the stars show that both nights were photometric to $\sim \pm$ 0.01 magnitudes.  Seeing was approximately 1.1\arcsec~FWHM.  Image composites formed by shifting the R-band images to a common center are shown in Figure (\ref{images}).  

Observations were also taken on UT 2017 October 28 using  the 3.5 meter diameter WIYN telescope, located at Kitt Peak National Observatory in Arizona.  We used the One Degree Imager (ODI) camera, a large CCD array which is mounted at the Nasmyth focus and has an image scale of 0.11\arcsec~pixel$^{-1}$ (Harbeck et al, 2014).  Observations were taken through the Sloan g' ($\lambda_c$ = 4750\AA, FWHM = 1500\AA), r' ($\lambda_c$ = 6250\AA, FWHM = 1400\AA) and i' filters ($\lambda_c$ = 7500\AA, FWHM = 1250\AA).  Seeing was variable in the range 0.8\arcsec~to 1.2\arcsec. Non-sidereal guiding with ODI is implemented through a guide window in the focal plane and is limited by the window size. We obtained 40 red filter integrations each of 180 s, each resulting in stellar images trailed by $\sim 9.5$\arcsec. Data reduction was performed using the Quickreduce pipeline (Kotulla 2014), and photometric calibration of the data was made with reference to the Pan STARRS system (Magnier et al.~2016).  Subsequent transformation from the Sloan filters to BVR filters employed at the NOT was made according to Jordi et al.~(2006).

\subsection{Photometry}
We used two-aperture photometry to measure the NOT images.   By experimentation, we selected a photometry aperture 10 pixels (2.14\arcsec) in radius to measure U1, but scaled the measurement to a 30 pixel (6.42\arcsec) radius aperture using measurements of the standard stars to estimate the flux between the two.  Sky subtraction was determined from the median signal in a concentric annulus extending from 6.42\arcsec~to 17.12\arcsec, for both U1 and the standard stars.  Two-aperture photometry is useful when, as here, the target is faint  because it reduces the uncertainty introduced by the sky noise.    

The apparent magnitudes are listed in Table (\ref{photometry}).  U1 showed strong brightness variations on each night of observation, as well as progressive fading between nights due to the changing observational geometry (Table \ref{geometry}).  We computed absolute magnitudes using 

\begin{equation}
H_{\lambda} = m_{\lambda} - 5\log_{10}(r_H \Delta) - \Phi(\alpha)
\label{absolute}
\end{equation}

\noindent where $r_H$ and $\Delta$ are the heliocentric and geocentric distances expressed in AU, and $\Phi(\alpha)$ is the phase function at phase angle $\alpha$.  We used $\Phi(\alpha) = \beta \alpha$, with $\beta$ =  0.04 magnitudes degree$^{-1}$.  This linear darkening function neglects brightening due to opposition surge but is suggested by observations of numerous low albedo asteroids at phase angles comparable to those of U1.  It should provide a useful correction for phase effects over the limited range of angles swept by U1 in our data.  

From the absolute magnitude we estimate the effective scattering cross-section, $C_e$ [m$^2$], using

\begin{equation}
C_e = \frac{1.5\times10^{12}}{p_V}10^{-0.4 H_V}
\label{C_e}
\end{equation}

\noindent in which $p_V$ is the V-band geometric albedo and $H_V$ is the absolute V-band magnitude.  The albedo of U1 is observationally unconstrained.  However the albedos of most solar system  bodies are within   a factor of three of $p_V$ = 0.1, so we adopt this value.  

\begin{figure}
\epsscale{.99}

\plotone{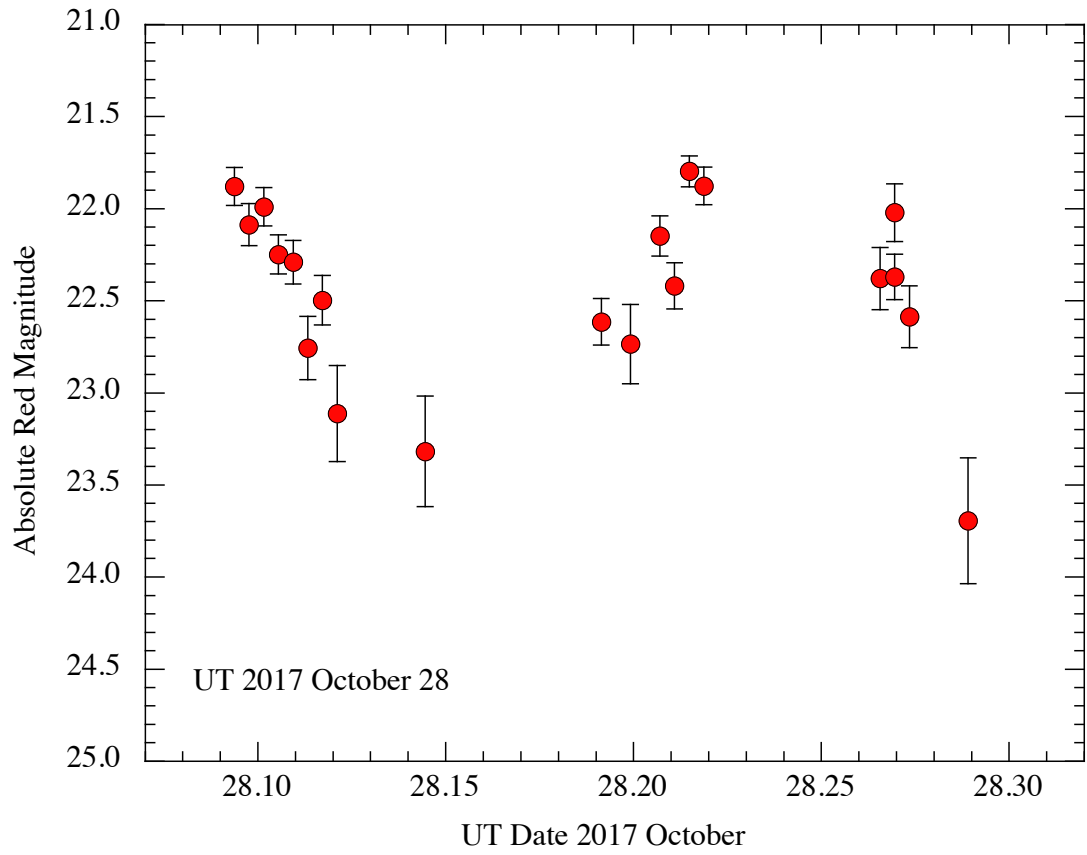}
\caption{Lightcurve from the WIYN telescope on UT 2017 October 28.     \label{oct28_lightcurve}}
\end{figure}

In the WIYN data,  the average absolute red magnitude is $H_R \sim$  22.5 (Figure \ref{oct28_lightcurve}).  With V-R = 0.45 and using Equation (\ref{C_e}), the mean magnitude corresponds to $C_e = 9.9 \times 10^{3}$ m$^2$ and to an equal-area circle of radius $(C_e/\pi)^{1/2}$ = 55 m. With this radius and nominal density $\rho$ = 10$^3$ kg m$^{-3}$, the approximate mass of U1 is $\sim$10$^9$ kg.

\subsection{Lightcurve}
Variations in $H_R$ (Table (\ref{photometry}) and Figure (\ref{oct28_lightcurve})) strongly suggest rotation of an irregular body.   The absolute magnitude varies in the range $21.5 \lesssim H_R \lesssim 23.5$, corresponding to cross-sections 4$\times10^{3} \lesssim C_e \lesssim 25\times10^{3}$ m$^2$ by Equation (\ref{C_e}).  Interpreted as a shape effect due to rotation of an $a \times a \times b$ ellipsoid about minor axis $a$, we find $b$ = 230 m and $a$ = 35 m.  The implied $\sim$6:1 axis ratio is  a lower limit because of the effects of projection, and is extreme relative to  most asteroids  (c.f.~Masiero et al.~2009, Dermawan et al.~2011, Hergenrother and Whiteley 2011).  However, we note that the 5 km diameter asteroid 4116 (Elachi) shows a range of 1.6 magnitudes ($b/a$ = 4.3:1) that is almost as large (Warner and Harris 2011).  The lightcuve of U1 might result from shape and albedo variations combined. 

\begin{figure}
\epsscale{.99}

\plotone{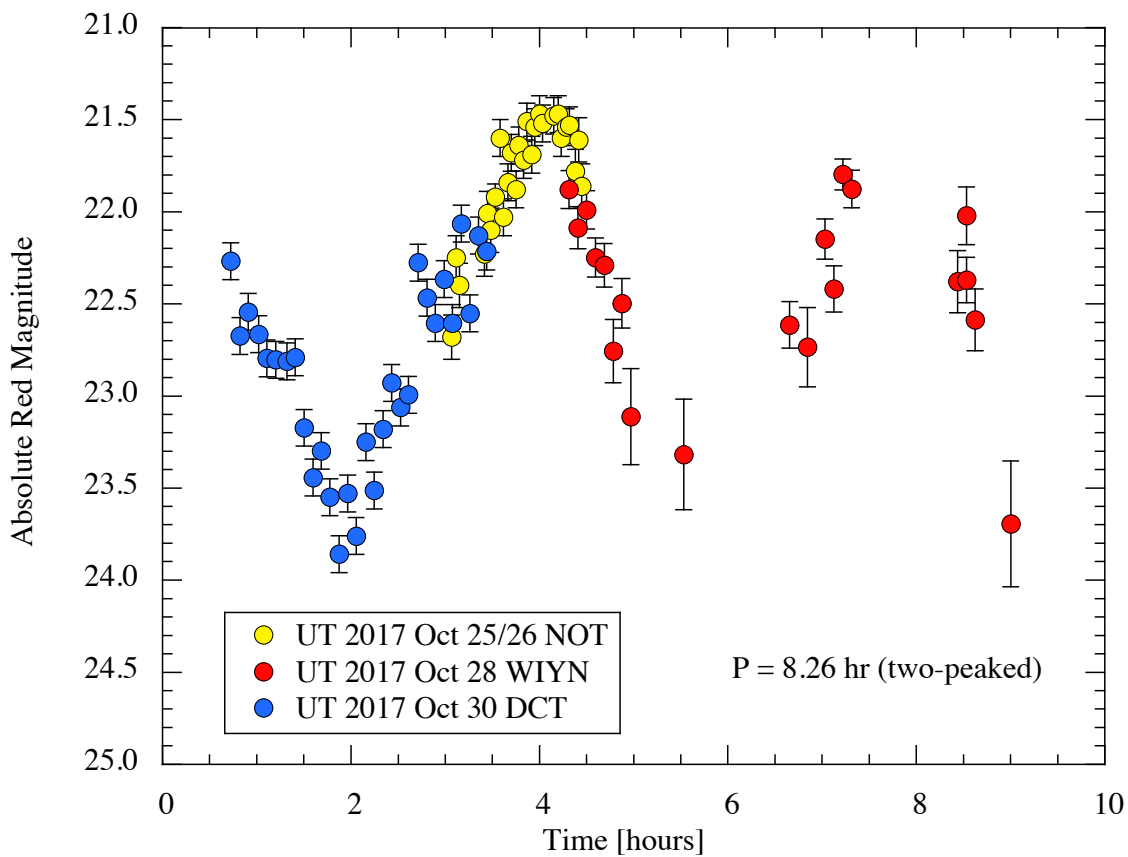}
\caption{Sample lightcurve phased to the two-peaked period $P$ = 8.26 hr using data from Table (\ref{photometry}) and from the Discovery Channel Telescope (DCT; Knight et al.~2017). The derived period is non-unique because of aliasing.  Time is plotted in hours, with an arbitrary zero point.  No correction has been made for the changing phase-angle bisector.  The October 28 data have been slightly vertically offset in order to improve the phasing, by an amount consistent with the phase function uncertainty.  \label{phased_0230}}
\end{figure}

We used phase dispersion minimization to estimate the possible rotational periods of U1.  We combine the data from Table (\ref{photometry}) with measurements from UT October 30 reported by Knight et al.~(2017), which we digitized and reduced to absolute magnitudes.  No unique solution was found, owing to aliasing in the data caused by incomplete temporal coverage. One of the possible lightcurve periods, $P$ = 8.256 hr, is shown for illustration in Figure (\ref{phased_0230}).  We expect that the addition of photometry from other telescopes will be used to improve the accuracy of the rotational period.   However, finding the exact period is  scientifically less important than the observation that the period of U1 falls squarely within the range of values observed for small asteroids in our solar system (Masiero et al.~2009).   

Approximating the shape of U1 as a strengthless $a \times a \times b$ ellipsoid with $b > a$ and in rotation about a minor axis, we can estimate the density needed to retain material against loss to centripetal forces.  We find

\begin{equation}
\rho = \frac{3\pi}{G P^2} \left(\frac{b}{a}\right)^2
\label{rho}
\end{equation}

\noindent where $G$ is the gravitational constant.  With $b/a \sim$ 6 and $P$ = 8.3 hr, we obtain $\rho \sim$ 6,000 kg m$^{-3}$, an implausibly high density that means only that U1 has non-negligible cohesive strength.

\subsection{Colors}
\begin{figure}
\epsscale{.9}

\plotone{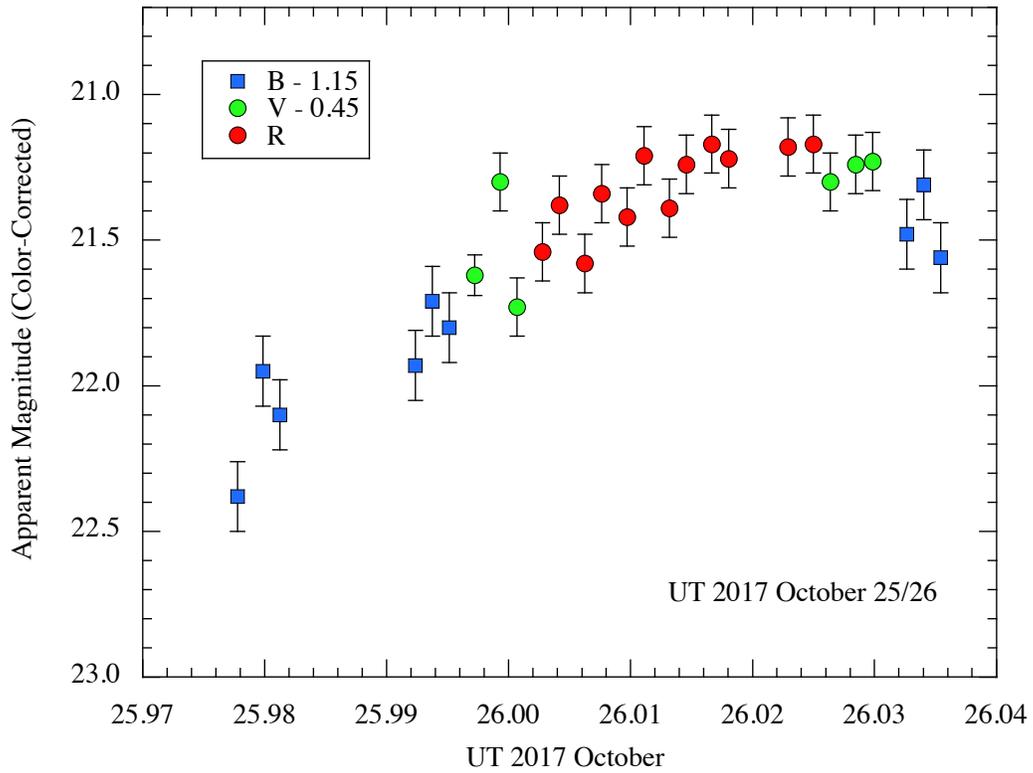}
\caption{Lightcurve from NOT data taken UT 2017 October 26.  Data from the BVR filters are color-coded as shown.  The B and V data have been shifted vertically in the plot  in order to estimate the B-V and V-R colors while retaining the shape of the rotational lightcurve.  \label{oct26_lightcurve}}
\end{figure}

Unfortunately, the color data at NOT were  interleaved too slowly to follow the lightcurve variations, so that substantial corrections for the rotational variation of the scattered light must be made when computing the B-V and V-R color indices.  For this purpose, we plotted the multi-filter color data as a series and applied vertical adjustments to the measurements in order to produce a smooth lightcurve (Figure \ref{oct26_lightcurve}).  This procedure is valid provided the color of U1 is constant with respect to rotation, as is true for most asteroids. The resulting colors are 

\begin{equation}
B-V = 0.70\pm0.06,  ~~V-R = 0.45\pm0.05.
\label{colors}
\end{equation}

\begin{figure}
\epsscale{.8}

\plotone{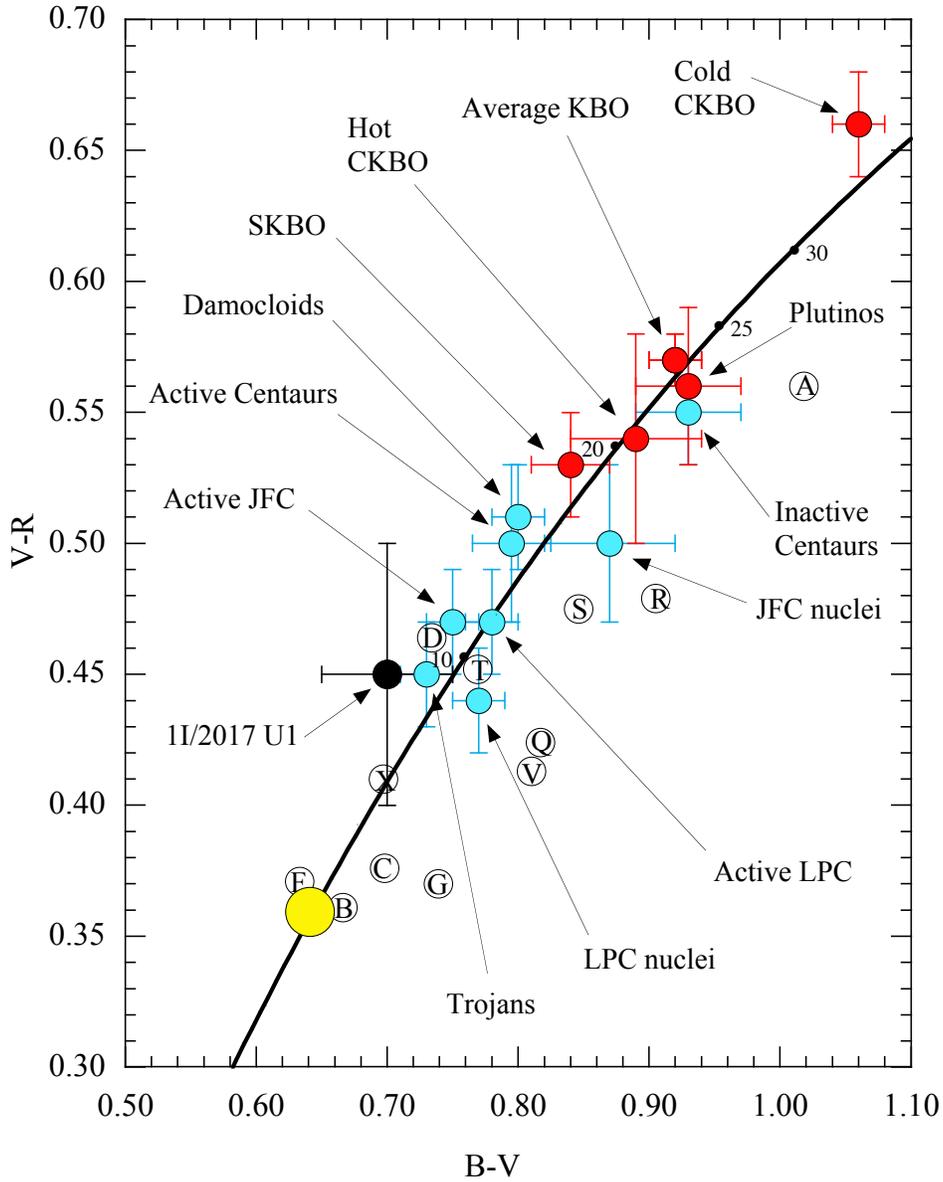}

\caption{B-V vs.~V-R color-color plot adapted from Jewitt (2015) to show the location of U1 relative to other solar system small-body groups.  The red circles denote sub-populations of the Kuiper belt, the blue circles show the mean colors of inner solar system populations.  The large yellow circle marks the color of the Sun. Letters show asteroid spectral types according to Dandy et al. (2003).  Error bars on U1 are $\pm$1$\sigma$ from the NOT data.  All other error bars are the 1$\sigma$ errors on the means of many measurements per population.   \label{colorcolor}}
\end{figure}

\noindent On a color-color diagram of the small-body populations of the solar system  (Figure \ref{colorcolor}) the optical colors of U1 are seen to be similar to the colors of the D-type Jovian Trojans  and several other inner-solar system groups, including the nuclei of both short-period (Kuiper belt) and long-period (Oort cloud) comets (Jewitt 2015).  On the other hand, the colors are quite different from the ultrared matter that is abundant in the outer solar system.  Specifically, the ultrared matter has, by definition (Jewitt 2002), a spectral slope $S' >$ 25\%/1000\AA~corresponding to B-R $>$ 1.60, whereas the color of U1 is B-R = 1.15$\pm$0.05.     The  normalized optical reflectivity gradient of U1 has been measured from spectra, albeit with very large uncertainties,  as $S'$ = 30$\pm$15\%/1000\AA~(Masiero 2017) and $S'$ = 10$\pm$6\%/1000\AA~(Ye et al.~2017).  

\subsection{Image Profile}
\begin{figure}
\epsscale{1.0}

\plotone{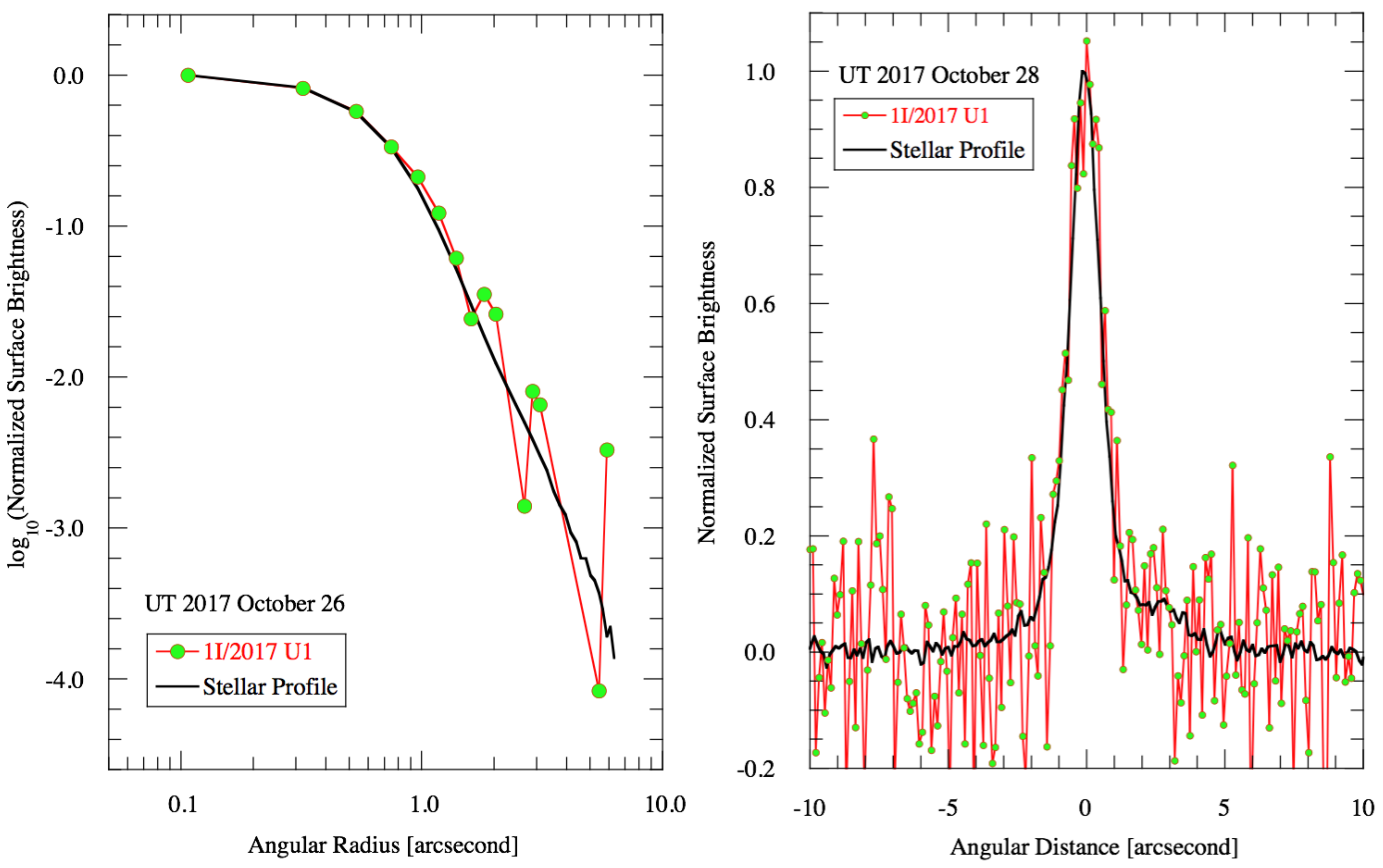}
\caption{(left) Surface brightness profile from NOT data taken UT 2017 October 26.  The circularly-averaged profile of 1I/2017 U1 (black line) is compared with the profile of field stars (red line)  in observations taken with sidereal tracking. (right) The surface brightness profile  on UT 2017 October 28 using the WIYN telescope and measured perpendicular to trailed field stars.  In neither case is a coma evident. \label{sbprofile}}
\end{figure}

The left-hand panel of Figure (\ref{sbprofile}) compares the surface brightness profile of U1 (combined integration time 1440 s through the R filter at NOT), with the surface brightness profile of field stars obtained from sidereally-guided images.  The surface brightness of U1 is normalized in the plot to a central value 23.8 magnitudes arcsecond$^{-1}$.  The right-hand panel shows the profile on UT 2017 October 28 from WIYN.  This profile is measured along a line perpendicular to the direction of field star trail, and averaged over the width of the trail.  In both cases the agreement between the profiles is remarkably good, with no evidence for a coma.

\section{Discussion}

\subsection{Comet or Asteroid?}
In order to place a limit to the outgassing of a coma implied by the non-detection of extended surface brightness, we need to know the morphology of the ejected material in the sky plane.  If the coma is produced in steady-state then, by the equation of continuity, the surface brightness of ejected dust should fall inversely with the angular distance from the nucleus, $\theta$.  In this case, the relation between the total magnitude of the coma, $m_C$, and the surface brightness at angular distance $\theta$, $\Sigma(\theta)$, is (Jewitt and Danielson 1984)

\begin{equation}
m_C = \Sigma(\theta) - 2.5\log \left(2\pi \theta^2\right)
\label{JD84}
\end{equation}

\noindent  From the profile (Figure \ref{sbprofile}) we determine that, at $\theta$ = 2\arcsec, a systematic brightening of the PSF by $\sim$1\% of the peak surface brightness would be detectable, corresponding to $\Sigma(2\arcsec)$ = 28.8 magnitudes arcsecond$^{-2}$.  Substituting into Equation (\ref{JD84}) gives the integrated magnitude of such a coma as $m_C$ = 25.3, which is fainter that the measured median R-band magnitude on this date, $m_R$ = 21.4, by $\delta m = m_C - m_R$ = 3.9 magnitudes. Therefore, a steady-state coma can carry no more than 10$^{-0.4\delta m} \sim$ 3\% of the total cross-section of U1.  With average $C_e \sim$ 10$^{4}$ m$^2$, the upper limit to the dust cross-section is only $C_d <$ 300 m$^2$.

Interpretation of this limit to the dust cross-section is highly model dependent, with the major unknowns being the particle size and the particle velocity.  In natural power-law size distributions, the cross-section is usually dominated by particles with size comparable to the wavelength of observation.  Given the wavelengths of our observations, we assume a particle radius $\overline{a}$ = 10$^{-6}$ m. We determine an upper limit to the rate of mass loss  in micron-sized particles as follows.  We assume that the particles are well-coupled to the outflowing gas so that they leave at the sound speed appropriate to the local equilibrium temperature, $V_d \sim$ 500 m s$^{-1}$.  The mass of dust is $M_d \sim \rho \overline{a} C_d$, where $\rho$ = 10$^3$ kg m$^{-3}$ is the assumed dust density.  Substituting gives $M_d <$ 0.3 kg.  The linear distance corresponding to  2\arcsec~at geocentric distance $\Delta$ =  0.458 AU is $\ell$ = 6.7$\times$10$^5$ m and the time for a dust particle to cross this distance is $\tau = \ell/V_d \sim$  1400 s.  The order of magnitude production rate is then $dM_d/dt = M_d/\tau <$ 2$\times$10$^{-4}$ kg s$^{-1}$.  For comparison, we calculated $f_s$, the sublimation flux from an exposed, perfectly absorbing water ice surface oriented normal to the Sun and located at $r_H$ = 1.4 AU, finding $f_s$ = 2.3$\times$10$^{-4}$ kg m$^{-2}$ s$^{-1}$.  The limit to the near-nucleus dust in U1 thus sets a limit to exposed ice on the surface of the body of area $A = f_s^{-1} dM_d/dt \sim$ 1 m$^2$.  Expressed as the fraction of the projected surface area of the (assumed) ellipsoidal nucleus, this is $f_A = A/(\pi a b) \lesssim 10^{-5}$ and we emphasize that this is an upper limit because we have assumed the maximum speed (and minimum residence time) for the dust particles in the near-nucleus environment.  For comparison, the nuclei of weakly active cometary nuclei have $f_A \sim 10^{-2}$ (A'Hearn et al.~1995).  

The absence of measurable coma thus shows that the surface of U1 contains little ice.  However,  we cannot conclude from this that U1 is an asteroid.  The transport of heat by conduction in a solid is controlled by the thermal diffusivity, $\kappa$, equal to the ratio of the thermal conductivity to the product of the density and the specific heat capacity.  Common solid dielectric materials (rock, ice) have $\kappa \sim$10$^{-6}$ m$^2$ s$^{-1}$ but the porous materials found in the regoliths of asteroids and comets have much smaller values, $\kappa \sim$ 10$^{-8}$ m$^2$ s$^{-1}$ or even smaller.  The timescale to conduct heat over a distance, $d$, is given by $t_c = d^2/\kappa$.  U1 travelled from the orbit of Jupiter, where water ice sublimation typically begins, to discovery in about 8 months, or 2$\times$10$^7$ s.  Setting $t_c$ = 2$\times$10$^7$ s and $\kappa$ = 10$^{-8}$ m$^2$ s$^{-1}$, we estimate the thermal skin depth (at which the temperature is $\sim 1/e$ times the surface temperature), to be $d = (\kappa t_c)^{1/2} \sim$ 0.5 m.  Ice at substantially larger depths would remain close to the interstellar temperature $\sim$10 K,  impervious to the heat of the Sun even at perihelion (where the blackbody temperature is $\sim$ 560 K).  Therefore, it is not possible to state based on the current observations whether U1 is an asteroid or a comet, except in the purely observational sense dictated by the absence of measurable coma.  A meter-thick  mantle of involatile, cosmic ray-irradiated material is expected from long-duration exposure in the interstellar medium (Cooper 2003) and could explain the inactivity of U1.   It is even possible that slow inward propagation of perihelion heat will activate buried ice some time in the future, as has been observed in some outbound comets (e.g.~Prialnik and Bar-Nun 1992).  For this reason, we encourage continued observations of U1 as it leaves the solar system.

\subsection{Statistics}

\begin{figure}
\epsscale{.8}

\plotone{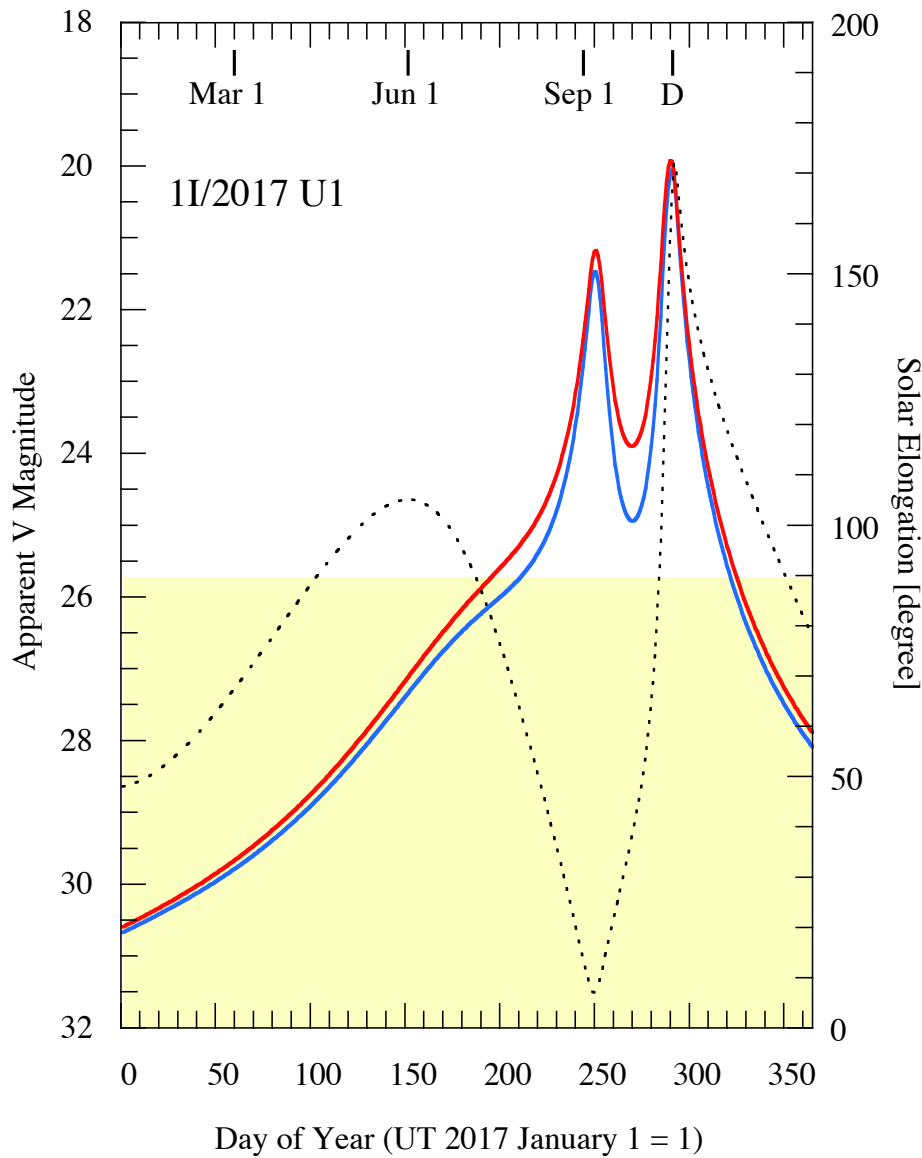}
\caption{Evolution of the apparent magnitude (red and blue curves, scale on the left-hand axis) and solar elongation (dotted line, scale on the right-hand axis) for 1I/2017 U1.  Red and blue curves show the magnitude assuming phase functions 0.03 magnitude degree$^{-1}$ and 0.04 magnitudes degree$^{-1}$, respectively.  The yellow-shaded region shows elongations, $\epsilon \le$ 90\degr, where ground-based observations are difficult.  The date of discovery is marked ``D''. \label{magplot}}
\end{figure}

Lastly, in Figure (\ref{magplot}) we show the apparent magnitude of U1 as a function of time in 2017, computed from Equation (\ref{absolute}) with $H_V$ = 22.95. Assumed phase functions $\beta$ = 0.03 and 0.04 magnitudes degree$^{-1}$ are shown as solid red and blue lines, respectively, while the solar elongation is indicated by a black dotted line.  It is obvious from the figure that U1 escaped detection for most of the year by being too faint but also, when brighter and near perihelion in early September, by appearing too close to the Sun.  Detection (marked ``D'' in the figure) occurred at peak brightness as a result of passing within $\Delta \sim$ 0.4 AU of the Earth.  

U1 is a very small object fortuitously detected  in a magnitude-limited survey because it passed  close to  the Earth (Figure \ref{magplot}).  Consider that the apparent brightness of an object in reflected light, all-else considered equal, depends on the product $a^2 \Delta^{-2}$, where $a$ is the object radius.  An object 10x larger than U1 would be equally bright at 10x the geocentric distance and could be detected in a magnitude-limited survey, of equal sensitivity, within a volume $\Delta^3 = 10^3$ times larger.  We represent the size distribution of interstellar objects, per unit volume, by a power law such that the number of objects with radii between $a$ and $a + da$ follows $N_1(a) da = \Gamma a^{-q} da$, with $\Gamma$ and $q$ constant.  Then the number observationally accessible in a magnitude-limited survey should scale as $\int N_1(a) a^3 da \propto a^{4-q}$.  The small size of the first detected interstellar object is an indication (subject to the limitations of the statistics of one) that the size distribution index is steep ($q \gtrsim 4$).

It is interesting to consider the implications of U1 for the statistics of  interstellar objects in the solar system.  The rate of detections of U1-like objects is $S = N_1 \pi (r_1 + \Delta)^2 F v_{\infty} \Psi$, where $N_1$ is the number of objects per unit volume, $r_1$ = 1 AU is the radius of the Earth's orbit, $F$ is the gravitational focusing factor by which orbits of interstellar bodies are concentrated owing to the Sun's gravity,  $v_{\infty}$ is the velocity  at infinity and $\Psi$ is the probability of passing within a distance $\Delta$ of Earth. The gravitational focusing factor is $F = (1 + v_e^2/v_{\infty}^2)$, where $v_e$ is the escape velocity at $r_1$.  We take $v_e$ = 42 km s$^{-1}$, $v_{\infty}$ = 25 km s$^{-1}$ to find $F$ = 4.  Given a random distribution of entering objects,  $\Psi \sim \Delta^2/(r_1 + \Delta)^2$ and, with $\Delta$ = 0.4 AU, we find $\Psi \sim$ 0.1.  The detection of U1 by Pan STARRS (which has been operating at full efficiency for moving object detections for only a year or two)  implies a rate $S \sim$ 0.5 to 1 year$^{-1}$.  Solving for the density of objects, we find $N_1 \sim$ 0.1 AU$^{-3}$, and $\sim$10$^{15}$ pc$^{-3}$.    The latter number exceeds even the highest pre-detection estimates by an order of magnitude (Moro-Martin et al.~2009).  From $N_1$ we estimate that, at any one time, there are $\sim$ 10$^4$ interstellar bodies of U1-size closer to the Sun than Neptune. Each takes $\sim$10 years to cross the planetary region before returning to interstellar space.

%

\section{Summary}
We present observations of the interstellar object 1I/2017 U1.  We find that:

\begin{enumerate}
\item The optical colors, B-V = 0.70$\pm$0.06 and V-R = 0.45$\pm$0.05, overlap the mean colors of D-type Trojan asteroids and other, inner-solar system populations. They are inconsistent with the ultrared matter found in the Kuiper belt.

\item The lightcurve shows a mean absolute magnitude  $H_V$ = 22.95$\pm$0.10 with a range 2.0$\pm$0.2 magnitudes.  The mean magnitude corresponds to an equal area circle of radius $\sim$55 m, assuming geometric albedo $p_V$ = 0.1.  If due to shape, the range implies an unusually large axis ratio $\gtrsim$6:1, suggesting an ellipsoidal body with  semi-axes 230 $\times$ 35 m (projected into the plane of the sky).  

\item The lightcurve is consistent with a two-peaked period of $\sim$8.256 hr, but the period is not definitively determined as a result of aliasing in the data.    This period is unremarkable relative to the periods of similarly-sized solar system small-bodies.

\item No coma is detected,  setting a limit to the rate of loss of micron-sized dust particles $\lesssim$2$\times$10$^{-4}$ kg s$^{-1}$.  Water ice covering a few m$^2$ of the surface would sublimate at this or a larger rate, in thermal equilibrium with sunlight.  The ice-covered fraction of the surface is $\lesssim$10$^{-5}$, some 10$^2$ or 10$^3$ times smaller than is typical for the nuclei Jupiter family comets.

\item The thermal conduction skin depth for U1's  8-month plunge through the inner solar system is only $\sim$0.5 m.  Ice could survive at near-interstellar temperatures beneath a thin refractory mantle, perhaps consisting of material rendered involatile by prolonged exposure to cosmic rays in the interstellar environment.

\item The number density of U1-like interstellar objects is $\sim$0.1 AU$^{-3}$, meaning that $\sim$10$^4$ such objects exist closer to the Sun than Neptune.
\end{enumerate}

\acknowledgments

We thank Jing Li, Pedro Lacerda, Joe Masiero and Quan-Zhi Ye for reading the manuscript and the anonymous referee for a rapid response. Based in part on observations made with the Nordic Optical Telescope, operated by the Nordic Optical Telescope Scientific Association at the Observatorio del Roque de los Muchachos, La Palma, Spain, of the Instituto de Astrofisica de Canarias.  We thank Daniel Harbeck for his vital contribution to making the ODI observations possible and Pere Blay, Amanda Djupvik, Sune Dyrbye, Francisco
Galindo-Guil, Tapio Pursimo, Ana Sagues and John Telting for help with the observations at the Nordic Optical Telescope. Based in part on observations at Kitt Peak National Observatory, National Optical Astronomy Observatory, which is operated by the Association of Universities for Research in Astronomy (AURA) under a cooperative agreement with the National Science Foundation. 



{\it Facilities:}  \facility{NOT, WIYN}.




\clearpage








\clearpage

\begin{deluxetable}{lllllllllr}
\tabletypesize{\scriptsize}
\tablecaption{Journal of Observations
\label{geometry}}
\tablewidth{0pt}
\tablehead{ \colhead{UT Date } & Telescope\tablenotemark{a}      & \colhead{$r_H$\tablenotemark{b}}  & \colhead{$\Delta$\tablenotemark{c}} & \colhead{$\alpha$\tablenotemark{d}} & \colhead{$\theta_{-\odot}$\tablenotemark{e}} & \colhead{$\theta_{-V}$\tablenotemark{f}} &  Conditions   }
\startdata

2017 Oct 25/26  & NOT & 1.384 &  0.429 & 20.7 & 73.3 & 184.8 & Photometric, 1.1\arcsec~seeing\\
2017 Oct 28 & WIYN & 1.436 & 0.502 & 23.1 & 74.2 & 193.5 & Photometric, 0.8 - 1.2\arcsec~seeing\\
2017 Oct 29/30 & NOT & 1.479 & 0.565 & 24.5 & 74.5  &  199.2 & Photometric, 1.0\arcsec~seeing\\

\enddata


\tablenotetext{a}{NOT = 2.5 m Nordic Optical Telescope, WIYN = 3.5 m Wisconsin-Indiana-Missouri-NOAO Telescope}
\tablenotetext{b}{Heliocentric distance, in AU}
\tablenotetext{c}{Geocentric distance, in AU}
\tablenotetext{d}{Phase angle, in degrees}
\tablenotetext{e}{Position angle of the projected antisolar direction, in degrees }
\tablenotetext{f}{Position angle of the negative heliocentric velocity vector, in degrees}

\end{deluxetable}

\clearpage

\begin{deluxetable}{lcclllllllr}
\tabletypesize{\scriptsize}
\tablecaption{Photometry
\label{photometry}}
\tablewidth{0pt}
\tablehead{ \colhead{UT Date\tablenotemark{a} } & Telescope\tablenotemark{b}      & \colhead{FILT\tablenotemark{c}}  & \colhead{$m_{\lambda}$\tablenotemark{d}} & \colhead{$H_{\lambda}$\tablenotemark{e}}    }
\startdata

2017 Oct 25.9778  &  NOT & B & 23.53 $\pm$ 0.12 & 23.83 \\
2017 Oct 25.9799  &  NOT & B & 23.10 $\pm$ 0.12 &  23.40 \\
2017 Oct 25.9813  &  NOT & B & 23.25 $\pm$ 0.12 &  23.55  \\
2017 Oct 25.9924  &  NOT & B & 23.08 $\pm$ 0.12  &  23.38 \\
2017 Oct 25.9937  &  NOT & B & 22.86 $\pm$ 0.12  &  23.16  \\
2017 Oct 25.9951  &  NOT & B & 22.95 $\pm$ 0.12  & 23.25   \\
2017 Oct 25.9972  &  NOT & V & 22.07 $\pm$ 0.07  &  22.37 \\
2017 Oct 25.9993  &  NOT & V & 21.75 $\pm$ 0.10  &  22.05 \\
2017 Oct 26.0007  &  NOT & V & 22.18 $\pm$ 0.10  &  22.48 \\
2017 Oct 26.0028  &  NOT & R & 21.54 $\pm$ 0.10  &  21.84  \\
2017 Oct 26.0042  &  NOT & R & 21.38 $\pm$ 0.10  & 21.68   \\
2017 Oct 26.0063  &  NOT & R & 21.58 $\pm$ 0.10  & 21.88  \\
2017 Oct 26.0076  &  NOT & R & 21.34 $\pm$ 0.10  & 21.64  \\
2017 Oct 26.0097  &  NOT & R & 21.42 $\pm$ 0.10  & 21.72  \\
2017 Oct 26.0111  &  NOT & R & 21.21 $\pm$ 0.10  &  21.51 \\
2017 Oct 26.0132  &  NOT & R & 21.39 $\pm$ 0.10  & 21.69  \\
2017 Oct 26.0146  &  NOT & R & 21.24 $\pm$ 0.10  &  21.54 \\
2017 Oct 26.0167  &  NOT & R & 21.17 $\pm$ 0.10  & 21.47  \\
2017 Oct 26.0181  &  NOT & R & 21.22 $\pm$ 0.10  & 21.52 \\
2017 Oct 26.0229  &  NOT & R & 21.18 $\pm$ 0.10  & 21.48 \\
2017 Oct 26.0250  &  NOT & R & 21.17 $\pm$ 0.10  & 21.47 \\
2017 Oct 26.0264  &  NOT & V & 21.75 $\pm$ 0.10  &  22.05   \\
2017 Oct 26.0285  &  NOT & V & 21.69 $\pm$ 0.10  & 21.99   \\
2017 Oct 26.0299  &  NOT & V & 21.68 $\pm$ 0.10  & 21.98  \\
2017 Oct 26.0326  &  NOT &  B & 22.63 $\pm$ 0.12  &  22.93   \\
2017 Oct 26.0340  &  NOT & B & 22.46 $\pm$ 0.12  & 22.76   \\
2017 Oct 26.0354  &  NOT &  B & 22.71 $\pm$ 0.12  & 23.01  \\

2017 Oct 28.0937	& WIYN &	R	&	22.19	$\pm$	0.14	&	21.98	\\
2017 Oct 28.0977	& WIYN &	R	&	22.40	$\pm$	0.16	&	22.19	\\
2017 Oct 28.1016	& WIYN &	R	&	22.30	$\pm$	0.14	&	22.09	\\
2017 Oct 28.1055	& WIYN &	R	&	22.56	$\pm$	0.17	&	22.35	\\
2017 Oct 28.1094	& WIYN &	R	&	22.60	$\pm$	0.17	&	22.39	\\
2017 Oct 28.1133	& WIYN &	R	&	23.07	$\pm$	0.27	&	22.86	\\
2017 Oct 28.1172	& WIYN &	R	&	22.81	$\pm$	0.22	&	22.60	\\
2017 Oct 28.1211	& WIYN &	R	&	23.43	$\pm$	0.37	&	23.21	\\
2017 Oct 28.1445	& WIYN &	R	&	23.63	$\pm$	0.41	&	23.42	\\
2017 Oct 28.1914	& WIYN &	R	&	22.93	$\pm$	0.23	&	22.72	\\
2017 Oct 28.1992	& WIYN &	R	&	23.05	$\pm$	0.28	&	22.84	\\
2017 Oct 28.2070	& WIYN &	R	&	22.46	$\pm$	0.15	&	22.25	\\
2017 Oct 28.2109	& WIYN &	R	&	22.73	$\pm$	0.19	&	22.52	\\
2017 Oct 28.2148	& WIYN &	R	&	22.11	$\pm$	0.11	&	21.90	\\
2017 Oct 28.2187	& WIYN &	R	&	22.19	$\pm$	0.12	&	21.98	\\
2017 Oct 28.2656	& WIYN &	R	&	22.69	$\pm$	0.22	&	22.48	\\
2017 Oct 28.2695	& WIYN &	R	&	22.68	$\pm$	0.19	&	22.47	\\
2017 Oct 28.2734	& WIYN &	R	&	22.90	$\pm$	0.24	&	22.69	\\
2017 Oct 28.2891	& WIYN &	R	&	24.01	$\pm$	0.59	&	23.79	\\

2017 Oct 29.9715\tablenotemark{f} & NOT & R & 22.15 $\pm$ 0.15  &   21.56  \\
2017 Oct 30.0896\tablenotemark{f} & NOT & R & 22.49 $\pm$ 0.15  &   21.90 \\
2017 Oct 30.0958\tablenotemark{f} & NOT & R & 22.20 $\pm$ 0.15  &   21.61 \\

\enddata


\tablenotetext{a}{UT date of the start of each exposure}
\tablenotetext{b}{Telescope: NOT = Nordic Optical Telescope, WIYN = Wisconsin-Indiana-Missouri-NOAO telescope}
\tablenotetext{c}{Filter}
\tablenotetext{d}{Apparent magnitude in the Bessel system.  Measurements from the WIYN telescope have been corrected from the Sloan filter system using R = r - 0.21.}
\tablenotetext{e}{Absolute magnitude.  The statistical uncertainty on the absolute magnitude is the same as on the apparent magnitude.  An additional uncertainty due to the unknown phase function has been ignored.}
\tablenotetext{f}{Data from October 30  have not been used in estimating the period.}

\end{deluxetable}

\clearpage

\clearpage

\clearpage

\clearpage

\clearpage

%
%
%

\clearpage

\clearpage

\clearpage

\end{document}